\documentclass[aps,twocolumn,showpacs,prl,preprintnumbers,letterpaper]{revtex4}
\usepackage{times}
\usepackage{amsmath,amssymb}
\usepackage{enumerate}
\usepackage{color}

\usepackage{graphicx}

\newcommand{\be}{\begin{equation}}
\newcommand{\ee}{\end{equation}}
\newcommand{\ba}{\begin{align}}
\newcommand{\ea}{\end{align}}

\begin{document}

\preprint{KUNS-2397}

\title{Nonequilibrium Phase Transitions and a Nonequilibrium Critical Point from Anti-de Sitter Space and Conformal Field Theory Correspondence}

\author{Shin Nakamura}
\email[E-mail: ]{nakamura@ruby.scphys.kyoto-u.ac.jp}
\affiliation{Department of Physics, Kyoto University, Kyoto 606-8502, Japan}


\begin{abstract}
We find novel phase transitions and critical phenomena that occur only outside the linear-response regime of current-driven nonequilibrium states. We consider the strongly-interacting $(3+1)$-dimensional ${\cal N}=4$ large-$N_{c}$ $SU(N_{c})$ supersymmetric Yang-Mills theory with a single flavor of fundamental ${\cal N}=2$ hypermultiplet as a microscopic theory. We compute its nonlinear non-ballistic quark-charge conductivity by using the AdS/CFT correspondence. We find that the system exhibits a novel nonequilibrium first-order phase transition where the conductivity jumps and the sign of the differential conductivity flips at finite current density. A nonequilibrium critical point is discovered at the end point of the first-order regime. We propose a nonequilibrium steady-state analogue of thermodynamic potential in terms of the gravity-dual theory in order to define the transition point. Nonequilibrium analogues of critical exponents are proposed as well. The critical behavior of the conductivity is numerically confirmed on the basis of these proposals. The present work provides a new example of nonequilibrium phase transitions and nonequilibrium critical points.
  
\end{abstract}

\pacs{05.70.Ln, 11.25.Tq, 05.70.Jk}

\maketitle

{\it Introduction}.---
Nonequilibrium physics is one of the central subjects in modern physics. Although the linear response theory provides a computational framework of the transport coefficients at the vicinity of thermal equilibrium, its extension to the nonlinear regime is still a great challenge. A key question in nonequilibrium physics is how to extract the macroscopic physics from the underlying microscopic theories when the systems are out of equilibrium.   

Recently, a great deal of attention has been paid to the anti-de Sitter space (AdS)/conformal field theory (CFT) correspondence~\cite{Maldacena:1997re,Gubser:1998bc}.
AdS/CFT correspondence is a map between a strongly interacting quantum gauge theory and a classical gravity. The map is essentialy established at the level of microscopic theory. However, it also provides a new picture for statistical mechanics. 
The macroscopic physics, such as thermodynamics, of the gauge particles appears naturally in terms of the black hole physics on the gravidy side~\cite{Witten:1998zw}.
The point is that the coarse graining of the gauge theory is accomplished in the gravity side just by solving the classical equations of motion.
This simplification tempts us to apply AdS/CFT to many-body physics out of equilibrium.

Usually, a drawback of AdS/CFT is that we need to deal with an idealized gauge theory that is not exactly realized in nature at the microscopic level, in order to ensure that the correspondence is well defined.
However, the aim of nonequilibrium statistical mechanics is to describe macroscopic properties common to a wide range of many-body systems regardless of the details of each microscopic theory. Our ultimate goal is to obtain new information on such macroscopic physics that is shared by the actual systems in nature.
Phase transition and critical phenomena are ideal places to look for such information in light of their possible universality.

In this Letter, we study nonequilibrium phase transitions~\cite{nePT} 
by using the AdS/CFT correspondence. 
Our system consists of a strongly interacting gauge-theory plasma driven to the
nonequilibrium steady state (NESS) by a constant current. We discover novel nonequilibrium phase transitions and a nonequilibrium critical point in the nonlinear regime. We develop a formalism to analyze the nonequilibrium phase transitions as well: we propose NESS analogues of thermodynamic potential and critical exponents.

Our phase transitions are associated with the nonlinear conductivity of our system. It has been shown in Ref.~\cite{Nakamura:2010zd} that the gauge-theory plasma, which will be defined in detail later, exhibits negative differential conductivity (NDC) in low-current-density regions. This NDC is categorized as S-shaped NDC (SNDC) in Ref.~\cite{book}. Here, the differential conductivity is defined as $\partial J/\partial E$, where $J$ and $E$ are the current density and the external electric field acting on the charge carriers, respectively. The difference from the conventional conductivity $\sigma=J/E$ is that the differential conductivity here can be either negative or positive, whereas $\sigma$ cannot be negative. In fact, SNDC has been experimentally observed in various current-driven systems~\cite{book}, including in systems of strongly correlated electrons (for example, see Ref.~\cite{Oka-Aoki}). 
It has been shown in Ref.~\cite{Nakamura:2010zd} that the NDC is converted to positive differential conductivity (PDC) in the high current-density regions. The transition from NDC to PDC was observed to be smooth in Ref.~\cite{Nakamura:2010zd}: it is a crossover.

In this Letter, we discover first-order and second-order phase transitions between NDC and PDC. We also observe critical phenomena at the critical point.
As far as the author knows, the current-driven nonequilibrium phase transitions between NDC and PDC, and the associated critical point, have been reported neither experimentally nor theoretically so far: the present work provides a new model of nonequilibrium phase transitions and nonequilibrium critical points.

{\it Microscopic theory}.---
We choose our microscopic theory by asking how much of its AdS/CFT duality is established, rather than asking whether the microscopic details are realized in nature, since we are interested in only the macroscopic properties of NESS that may be independent of the microscopic details.
One of the gauge theories whose AdS/CFT duality is well-established is the $(3+1)$-dimensional strongly-coupled $SU(N_{c})$ ${\cal N}=4$ super-symmetric Yang-Mills theory (${\cal N}=4$ SYM) at the large-$N_{c}$ limit with a single flavor of fundamental ${\cal N}=2$ hypermultiplet. This is a supersymmetric cousin of quantum chromodynamics (QCD), but its supersymmetry is broken at finite temperatures. 
The ${\cal N}=4$ SYM sector (gluon sector) contains the gauge particles in the adjoint representation, which we call ``gluons''  in this Letter.
The ${\cal N}=2$ hypermultiplet sector (quark sector) contains particles in the fundamental representation and anti-fundamental representation; we call these particles ``quarks'' and ``antiquarks,'' respectively. The quark (antiquark) carries a unit of positive (negative) quark charge. These particles play the role of, for example, electrons and holes in condensed matter, that is, the role of charge carriers. The interaction among these particles is mediated by the gluons; the gluons play the role of phonons, for example.

We consider the conductivity associated with the quark charge, and it is the quark-charge current that drives the quark sector out of equilibrium. NESS is realized in the following manner~\cite{Karch:2008uy}. We set the large-$N_{c}$ limit that makes the degree of freedom (DOE) of the gluon sector [which is $O(N_{c}^{2})$] sufficiently larger than that of the quark sector [which is $O(N_{c})$]. We also set the gluon sector equilibrium at a definite temperature $T$. The interaction between the gluon sector and the quark sector generates a dissipation in the presence of the quark current: the gluons absorb the momentum and the energy of the charge carriers. Because of the large DOE, the heat capacity of the gluon sector is sufficiently large, and the temperature of the gluon sector is well approximated as a constant. The gluon sector plays the role of heat bath, and the NESS of the quark sector is realized by putting the dissipation into the heat bath and the work of the external field in balance. 

Let us specify the conditions we impose on the system.
We consider a neutral system where the total quark-charge density is zero. This means that the finite current is realized by equal numbers of quarks and antiquarks flowing in opposite directions. The system works as an insulator when $E$ and $T$ are sufficiently small compared to the mass of the quark, but a strong enough electric field will break the insulation~\cite{Erdmenger:2007bn,Albash:2007bq}. Our charge carriers are those pair created by the external field \cite{Karch:2007pd} in the insulation breaking.
We assume that the system is steady and homogeneous. We also assume that the system has an infinitely large volume:
note that we are {\em not} dealing with mesoscopic systems.
We consider current-driven phenomena, and choose $J$ (the quark-charge current density) as our control parameter.
In this sense, $E$, the external field acting on the quark charge, is taken as a function of $J$. We choose the electric field (and hence the current) to be in the $x$ direction.
We employ the natural units $c=\hbar=k_{\mbox{\scriptsize B}}=1$.

{\it Nonlinear conductivity in AdS/CFT}.---
The gravity dual of our microscopic theory is the so-called D3-D7 system~\cite{Karch:2002sh}. The computational technique of nonlinear conductivity has been proposed in \cite{Karch:2007pd} in the framework of the AdS/CFT correspondence, and we follow it. 
We sketch the idea of \cite{Karch:2007pd} below to define our notation, and one may consult Refs.~\cite{Karch:2002sh,Karch:2007pd} for more details.
Our proposals for thermodynamic potential and critical exponents shall be given later.
 
The gluon sector at finite temperature is mapped to the gravity theory on a curved geometry, which is a direct product of a five-dimensional AdS-Schwarzschild black hole (AdS-BH) and $S^{5}$~\cite{Witten:1998zw}. 
The metric of the AdS-BH part is given by~\cite{ft1}
\begin{eqnarray}
ds^{2}=
-\frac{1}{z^{2}}\frac{(1-z^{4}/z^{4}_{H})^{2}}{1+z^{4}/z^{4}_{H}}dt^{2}
+\frac{1+z^{4}/z^{4}_{H}}{z^{2}}d\vec{x}^{2}+\frac{dz^{2}}{z^{2}},
\label{AdS-BH}
\end{eqnarray}
where $z$ ($0\le z \le z_{H}$) is the radial coordinate of the black hole geometry. 
The AdS-BH has a horizon at $z=z_{H}$ and a boundary at $z=0$.  
The Hawking temperature, which corresponds to the temperature of the gluon sector (hence that of the heat bath), is given by $T=\sqrt{2}/(\pi z_{H})$;
$\vec{x}$ and $t$ denote the $(3+1)$-dimensional spacetime coordinates on which the gauge theory is defined.
The metric of the $S^{5}$ part is given by
\begin{eqnarray}
d\Omega_{5}^{2}=d\theta^{2}+\sin^{2}\theta d\psi^{2}+\cos^{2}\theta d\Omega_{3}^{2},
\end{eqnarray}
where $0\le \theta\le \pi/2$, and $d\Omega_{d}$ is the volume element of the unit $d$-dimensional sphere. The radius of the $S^{5}$ part is taken to be 1 for simplicity. This is equivalent to choosing the 't Hooft coupling $\lambda$ of the gauge theory at $\lambda=(2\pi)^{2}/2$. 

The quark sector is mapped to a D7-brane~\cite{Polchinski:1998rq}
embedded to the above geometry. The three-dimensional part of the D7-brane is wrapped on the $S^{3}$ part in $S^{5}$, the radius of which is given by $\cos\theta$:
the configuration of the D7-brane is specified by the function $\theta(z)$~\cite{Karch:2002sh}. 
We employ the probe approximation where the backreaction of the D7-brane to the AdS-BH is neglected. This is justified at the large-$N_{c}$ limit, and is consistent with the picture of the AdS-BH as the heat bath. 
The behavior of $\theta(z)$ at the vicinity of the boundary is related to the current quark mass $m_{q}$~\cite{Karch:2002sh} as
\begin{eqnarray}
\theta(z)=m_{q} z + \frac{1}{2}\left[\langle \bar{q} q \rangle/N+m_{q}^{3}/3 \right] z^{3}+ O(z^{5}).
\end{eqnarray}
Here, $\langle \bar{q} q \rangle$ denotes the chiral condensate~\cite{ft2}.
Throughout the analysis, we fix $m_{q}$ at a designed value.
There is a $U(1)$ gauge field $A_{\mu}$ on the D7-brane.
$E$ and $J$ are related to $A_{x}$ through 
\begin{eqnarray}
A_{x}(z,t)=-Et+{\rm const.}+J(2N)^{-1} z^{2}+O(z^{4}),
\end{eqnarray}
where $N=N_{c}/(2\pi)^{2}$ in our convention and we have employed the gauge $\partial_{x}A_{t}=0$ \cite{Karch:2007pd}. 
The dynamics of $\theta(z)$ and $A_{x}(z,t)$ is governed by the D7-brane action, 
$S_{\text D7}=\int dt d^{3}x dz {\cal L}_{\text D7}$, where ${\cal L}_{\text D7}$ is explicitly written as~\cite{Karch:2007pd}
\begin{eqnarray}
{\cal L}_{\text D7}=-N g_{xx} \cos^{3}\theta 
\sqrt{
|g_{tt}|g_{xx}g_{zz}
-g_{zz}(\dot{A}_{x})^{2}+|g_{tt}|(A^{\prime}_{x})^{2}
}.
\end{eqnarray}
Here, the prime (the dot) denotes the differentiation with respect to $z$ ($t$); 
$g_{tt}, g_{xx}$ and $g_{zz}$ are the components of the induced metric on the D7-brane, and they are equal to those of the metric of AdS-BH except for $g_{zz}=1/z^{2}+\theta^{\prime}(z)^{2}$. 
If we regard $z$ as a ``time coordinate,'' $J=\partial {\cal L}_{\text D7}/\partial A_{x}^{\prime}$ is a conserved ``canonical momentum,'' since ${\cal L}_{\text D7}$ does not contain $A_{x}$ explicitly. Then it is convenient to introduce a ``Routhian'' $\widetilde{\cal L}_{\text D7}={\cal L}_{\text D7}-A_{x}^{\prime}\partial {\cal L}_{\text D7}/\partial A_{x}^{\prime}$ ~\cite{Nakamura:2006xk,Kobayashi:2006sb}, which is given in terms of $\dot{A}_{x}$, $J$, $\theta$ and $\theta^{\prime}$ \cite{Karch:2007pd}:
\begin{eqnarray}
\widetilde{\cal L}_{\text D7}
=-
\sqrt{g_{zz}
(g_{xx}-\dot{A}_{x}^{2}/|g_{tt}|)(N^{2}|g_{tt}|g_{xx}^{2}\cos^{6}\theta-J^{2})
}.
\label{Routhian}
\end{eqnarray}
The Euler-Lagrange equation from (\ref{Routhian}) determines $\theta(z)$ under given $E=-\dot{A}_{x}$ and $J$.
The relationship between $E$ and $J$ is determined by requesting the on-shell D7-brane action to be real valued: the inside of the square root in (\ref{Routhian}) has to be positive semidefinite~\cite{Karch:2007pd}.
This condition yields 
\begin{eqnarray}
\sigma=N \: T (e^{2}+1)^{1/4}\cos^{3}\theta(z_{*}),
\end{eqnarray}
where $z_{*}=\left[\sqrt{e^{2}+1}-e\right]^{1/2} z_{H}$ and $e=2E/(\pi \sqrt{2\lambda}T^{2})$ \cite{Karch:2007pd};
$z=z_{*}$ is the location where the inside of the square root in (\ref{Routhian}) touches zero~\cite{Karch:2007pd}.

We need numerical analysis to obtain $\theta(z_{*})$ explicitly.
The boundary conditions we employ are $\theta(z)/z|_{z=0}=m_{q}$ and $\theta^{\prime}|_{z=z_{*}}=[B-\sqrt{B^{2}+C^{2}}]/(Cz_{*})$.
Here $B=3z_{H}^{8}+2z_{H}^{4}z_{*}^{4}+3z_{*}^{8}$ and $C=3(z_{*}^{8}-z_{H}^{8})\tan\theta(z_{*})$. The condition for $\theta^{\prime}|_{z=z_{*}}$ comes from the equation of motion (EOM) at $z=z_{*}$~\cite{Albash:2007bq,Nakamura:2010zd} with the assumption $\theta(z_{*})\neq\pi/2$. 
Note that $\theta^{\prime}|_{z=z_{*}}$ is given in terms of
$E$, $T$ and $J$ for given $\lambda$ and $N_{c}$. 
After the solution $\theta(z)$ is obtained, we estimate $m_{q}$ from $\theta(z)/z|_{z=0}$ as a function of $E$, $T$ and $J$. We choose $E$ (under given $T$ and $J$) so that $m_{q}$ agrees with the designed value.
Since the numerical analysis becomes unstable at $z=0$, $z=z_{*}$, and at $z=z_{H}$,
we avoid these points by introducing small cutoffs in the numerical computations.

\begin{figure}[htb]
  \begin{center}
    \includegraphics[keepaspectratio=true,height=70mm]{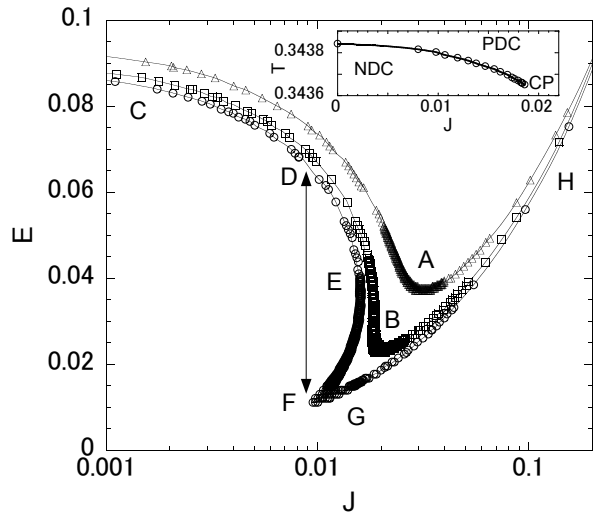}
  \end{center}
  \caption{The $J$-$E$ curves at $T=3.4379\times 10^{-1}> T_{c}$ (circle), $T=3.4365\times 10^{-1}=T_{c}$ (box), and $T=3.4337\times 10^{-1}<T_{c}$ (triangle). Inset: the phase diagram. CP stands for the critical point.}
  \label{fig:figure1-2.eps}
\end{figure}

{\it Nonequilibrium phase transitions}.---
Let us fix $N_{c}=40$ for numerical computation. We fix $m_{q}=1$ for simplicity.
The $J$-$E$ characteristics at various $T$ are shown in Fig.~\ref{fig:figure1-2.eps}. The system exhibits NDC in the small-$J$ region, whereas PDC is seen in the large-$J$ region. 
The NDC region is smoothly connected to the PDC region for $T<T_{c}$, showing a crossover at point A.
However, the curve for $T>T_{c}$ has an intermediate region (between points D and G), where three values of $E$ are possible at a given $J$. This multivalued nature of $E$ at a given $J$ has never been obtained in eariler works. In terms of the D7-brane dynamics, we have three different solutions to the EOM of $\theta(z)$. If we start in the small-$J$ region, $E$ has to jump to the lower value at somewhere in the intermediate region, and then PDC appears. Since $E$ (hence $\sigma$) changes discontinuously, we call it first-order transition. 
The boundary between the crossover regime and the first-order regime is found at $T=3.4365\times 10^{-1}=T_{c}$.
The differential resistivity $\partial E/\partial J$ diverges at $J=1.86 \times 10^{-2}=J_{c}$ (indicated by B) although $\sigma$ changes continuously. We call this transition at $(T,J)=(T_{c}, J_{c})$ a second-order transition. 
The minimum value of $E$ for each curve is the critical electric field for insulation breaking.

An immediate question is how to determine the transition point in the intermediate region. In equilibrium systems, the stable phase is the phase of minimum thermodynamic potential (TP), and the transition point is where the two phases share a common TP. However, the generalization of the idea of TP into nonequilibrium cases has not been completely established. 
One way to evaluate the NESS generalization of TP is to use the Maxwell construction~\cite{Bergman:2008sg}. However, this method works only for the pairs of conjugate variables. In our case, we need a TP as a function of $J$. Note that $J$ and $E$ are not conjugate to each other. The variable conjugate to $J$ is $A_{x}|_{z=0}$, which explicitly depends on time, and we cannot construct a time-independent TP by integrating $A_{x}|_{z=0}$ with respect to $J$: the method of Maxwell construction does not work for our purpose. 
Note that the Euclidean on-shell action of D7-brane cannot be employed as the TP in our system, since the reflection of real time is essential in nonequilibrium systems. If we insist with the Euclidean formalism, we encounter unremovable conical singularity in the gravity dual~\cite{Sonner:2012if}. In this sense, we do not employ the Euclidean action proposed in \cite{Alam:2012fw}. 
We need a new proposal for NESS generalization of TP within the Minkowski signature.

Interestingly, we can determine the most stable phase by considering the Hamiltonian of the D7-brane in the gravity dual. The question is which D7-brane configuration is most stable among the three possible solutions. 
Since the dynamics of the D7-brane in the dual picture is governed by classical mechanics, the most stable configuration is that of the smallest Hamiltonian.
We have seen that the D7-brane dynamics is governed by $\widetilde{\cal L}_{\text D7}$. Therefore, let us construct the Hamiltonian density 
\begin{eqnarray}
\widetilde{\cal H}_{\text D7}=\dot{A}_{x}\partial \widetilde{\cal L}_{\text D7}/\partial \dot{A}_{x}-\widetilde{\cal L}_{\text D7},
\end{eqnarray}
which is explicitly given by~\cite{ft3}
\begin{eqnarray}
\widetilde{\cal H}_{\text D7}
=g_{xx}\sqrt{|g_{tt}|g_{zz}}
\sqrt{\frac{N^{2}\cos^{6}\theta |g_{tt}|g_{xx}^{2}-J^{2}}{|g_{tt}|g_{xx}-E^{2}}}.
\label{Hamiltonian}
\end{eqnarray}
Note that $\widetilde{\cal H}_{\text D7}$ is {\em regular} at the horizon (where $g_{tt}=0$) and free from the IR divergence discussed in \cite{Karch:2008uy,Alam:2012fw}. 
The divergence at the horizon
in $\widetilde{\cal L}_{\text D7}$ is canceled by that in $\dot{A}_{x} \partial \widetilde{\cal L}_{\text D7}/\partial \dot{A}_{x}$ within the Legendre transformation. 
We propose to define the thermodynamic potential per unit 3d volume in our system by
\begin{eqnarray}
\widetilde{H}_{\text D7}(T,J;m_{q})=\lim_{\epsilon\to 0}\left [
\int^{z_{H}}_{\epsilon}dz \widetilde{\cal H}_{\text D7}-L_{\text{count}}(\epsilon)
\right],
\label{TP}
\end{eqnarray}
where $L_{\text{count}}$ denotes the counterterms that renormalize the divergence at the boundary $z=0$ (which corresponds to UV divergence in the gauge theory). $L_{\text{count}}$ is given by
\begin{eqnarray}
L_{\text{count}}=L_{1}+L_{2}-L_{F}+L_{f},
\end{eqnarray}
where $L_{1}$, $L_{2}$, $L_{F}$, and $L_{f}$ are explicitly given in \cite{Karch:2005ms,Karch:2007pd}. Note that the relative sign between $L_{F}$ and the others has to be flipped, compared to the counterterms for the action owing to the Legendre transformation.

It is found from our numerical analysis that the configurations between F and G have the smallest $\widetilde{H}_{\text D7}$ compared to those between E and F, and those between D and E at a given $J$.
Therefore, the transition point between NDC and PDC is D or F as is indicated by the arrow in Fig.~\ref{fig:figure1-2.eps}. Our system prefers the smallest dissipation under the current-controlled setup.

The next question is how to define the critical exponents.
We find that the NDC and PDC phases are connected via the crossover region on the phase diagram (shown in the inset of Fig.~\ref{fig:figure1-2.eps}): the symmetry of the system does not change through the transition.
This resembles the liquid-gas transitions and the Mott insulator-to-metal transitions in equilibrium systems, whose critical points are in the same universality class of the Ising model \cite{Papanikolaou}.
In the liquid-gas transitions, the critical exponents $\beta$ and $\delta$ are given by $\bigtriangleup \rho \propto |T-T_{c}|^\beta$ along the first-order transition line and $|\rho-\rho_{c}| \propto |P-P_{c}|^{1/\delta}$ along the $T=T_{c}$ line. 
Here, $\rho$ and $P$ are the density and the pressure, and $\rho_{c}$ and $P_{c}$ are their critical values.
$\bigtriangleup \rho$ is the difference of the density between the liquid phase and the gas phase. 
In the equilibrium Mott insulator-to-metal transitions, the critical exponents can be detected by using the conductivity instead of the density~\cite{Limelette,Papanikolaou}. Therefore, let us generalize the definition of the exponent $\beta$ into the nonequilibrium cases as 
\begin{eqnarray}
\bigtriangleup \sigma \propto |T-T_{c}|^{\beta},
\end{eqnarray}
where the temperature is that of the heat bath and $\bigtriangleup \sigma$ is the difference of the conductivity between the NDC phase and the PDC phase along the first-order transition line. We regard $\bigtriangleup \sigma$ as a probe of the order parameter. The pressure of the system is not a control parameter within the present setup, but we have $J$ instead. Let us define a new critical exponent $\tilde{\delta}$ by 
\begin{eqnarray}
|\sigma-\sigma_{c}| \propto |J-J_{c}|^{1/\tilde{\delta}},
\end{eqnarray}
along the $T=T_{c}$ line. 
\begin{figure}[htb]
  \begin{center}
    \includegraphics[keepaspectratio=true,height=70mm]{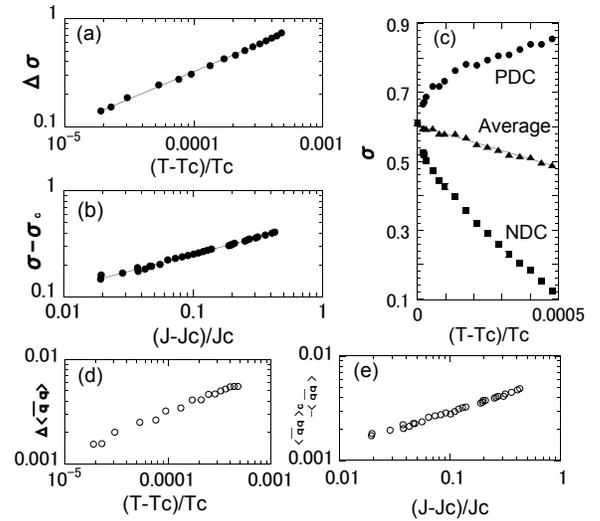}
  \end{center}
  \caption{Critical behaviors of various quantities. (a): $\bigtriangleup \sigma$, (b): $\sigma-\sigma_{c}$, (c): conductivities and the average, (d): $\bigtriangleup \langle \bar{q} q \rangle $, and (e): $\langle \bar{q} q \rangle_{c}-\langle \bar{q} q \rangle$.
  }%
  \label{fig:figure2.eps}
\end{figure}

The behaviors of the conductivity are plotted in Figs.~\ref{fig:figure2.eps}(a) and \ref{fig:figure2.eps}(b). Critical phenomena with $\beta=0.52\pm 0.03$ and $\tilde{\delta}=3.1\pm 0.2$ are numerically found.
The conductivities in the two phases and their average, along the first-order transition line, are shown in Fig.~\ref{fig:figure2.eps}(c). They resemble the coexistence line and the law of rectilinear diameter in the liquid-gas transitions~\cite{CL}: the average conductivity shows a linear behavior within the numerical error. 
In Figs.~\ref{fig:figure2.eps}(d) and \ref{fig:figure2.eps}(e), the behaviors of the chiral condensate are shown. The chiral condensate is more sensitive to the possible numerical errors since it is read by $\partial_{z}^{3}\theta|_{z=0}$: we need further analysis to estimate the precise values of the exponents. Currently, we observe preliminary values of the exponents $\beta_{\text{chiral}}\sim 0.4$ and $\tilde{\delta}_{\text{chiral}}\sim 3$, where we have defined the exponents by $\bigtriangleup \langle \bar{q} q \rangle \propto |T-T_{c}|^{\beta_{\text{chiral}}}$ along the first-order-transition line and by  $|\langle \bar{q} q \rangle-\langle \bar{q} q \rangle_{c}| \propto |J-J_{c}|^{1/\tilde{\delta}_{\text{chiral}}}$ along the $T=T_{c}$ line.
Here, $\bigtriangleup \langle \bar{q} q \rangle $ is the difference of $\langle \bar{q} q \rangle$ between the two phases and $\langle \bar{q} q \rangle_{c}$ is the chiral condensate at the critical point.


{\it Acknowledgments}:
The author thanks H. Fujii, H. Hayakawa, S. Inutsuka, T. Iritani, H. Kawai, Y. Hikida, S. Kinoshita, C. Maes, Y. Minami, S. Mukohyama, T. Oka, R. Sakano, S. Sasa, A. Shimizu, H. Suganuma, T. Takayanagi and H. Tasaki for useful comments. 
Discussions during the YITP conferences YITP-W-11-11, YITP-W-11-14, YITP-W-11-16, YITP-T-11-05 at the Yukawa Institute for Theoretical Physics at Kyoto University, the conferences KEK-TH2012, QMKEK4 at the High Energy Accelerator Research Organization (KEK), the GCOE symposiums in 2010 and in 2011 at Kyoto University, and the conference QHEC11 at Keio University, were useful. 
This work was supported in part by the Grant-in-Aid (GIA) for Scientific Research on Innovative Areas No. 2104, and GIA
for Challenging Exploratory Research No. 23654132.


\begin{thebibliography}{99}

\newcommand{\J}[4]{{\sl #1} {\bf #2} (#3) #4}
\newcommand{\andJ}[3]{{\bf #1} (#2) #3}
\newcommand{\AP}{Ann.\ Phys.\ (N.Y.)}
\newcommand{\MPL}{Mod.\ Phys.\ Lett.}
\newcommand{\NP}{Nucl.\ Phys.}
\newcommand{\PL}{Phys.\ Lett.}
\newcommand{\PR}{Phys.\ Rev.}
\newcommand{\PRL}{Phys.\ Rev.\ Lett.}
\newcommand{\ATMP}{Adv.\ Theor.\ Math.\ Phys.}
\newcommand{\JHEP}{JHEP}
\newcommand{\IJMP}{Int.\ J.\ Mod.\ Phys.}
\newcommand{\JETPL}{JETP\ Lett.}
\newcommand{\SJNP}{Sov.\ J.\ Nuc.\ Phys.}
\newcommand{\PTP}{Prog.\ Theor.\ Phys.}
\newcommand{\CMP}{Commun.\ Math.\ Phys.}


\bibitem{Maldacena:1997re}
  J.~M.~Maldacena,
  ``The large N limit of superconformal field theories and supergravity,''
  Adv.\ Theor.\ Math.\ Phys.\  {\bf 2}, 231 (1998),
  [Int.\ J.\ Theor.\ Phys.\  {\bf 38}, 1113 (1999)],
  [arXiv:hep-th/9711200].

\bibitem{Gubser:1998bc}
  S.~S.~Gubser, I.~R.~Klebanov and A.~M.~Polyakov,
  ``Gauge theory correlators from non-critical string theory,''
  Phys.\ Lett.\  B {\bf 428}, 105 (1998),
  [arXiv:hep-th/9802109];
%
  E.~Witten,
  ``Anti-de Sitter space and holography,''
  Adv.\ Theor.\ Math.\ Phys.\  {\bf 2}, 253 (1998),
  [arXiv:hep-th/9802150].

\bibitem{Witten:1998zw}
  E.~Witten,
  ``Anti-de Sitter space, thermal phase transition, and confinement in  gauge
  theories,''
  Adv.\ Theor.\ Math.\ Phys.\  {\bf 2}, 505 (1998),
  [arXiv:hep-th/9803131].

\bibitem{nePT}
For reviews of nonequilibrium phase transitions and nonequilibrium critical phenomena, see for example,
M.~Henkel, H.~Hinrichsen and S.~L\"{u}beck, {\sl Non-Equilibrium Phase Transitions. Vol. 1,} (Springer-Verlag, Dordrecht 2008);
M.~Henkel and M.~Pleimling, {\sl Non-Equilibrium Phase Transitions. Vol. 2,} (Springer-Verlag, Dordrecht 2010).

\bibitem{Nakamura:2010zd}
  S.~Nakamura,
  ``Negative Differential Resistivity from Holography,''
  Prog.\ Theor.\ Phys.\  {\bf 124}, 1105 (2010), 
  [arXiv:1006.4105 [hep-th]].

\bibitem{book}
E. Sch\"{o}ll, {\sl Nonlinear Spatio-Temporal Dynamics and
Chaos in Semiconductors} (Cambridge Univ. Press,  Cambridge 2001).

\bibitem{Oka-Aoki}
T. Oka and H. Aoki,
``Nonequilibrium Quantum Breakdown in a Strongly Correlated Electron System,''
Lect. Notes Phys. {\bf 762}, 251 (2009),
[arXiv:0803.0422 [cond-mat.str-el]].

\bibitem{Karch:2008uy}
  A.~Karch, A.~O'Bannon and E.~Thompson,
  ``The Stress-Energy Tensor of Flavor Fields from AdS/CFT,''
  JHEP {\bf 0904}, 021 (2009),
  [arXiv:0812.3629 [hep-th]].

\bibitem{Albash:2007bq}
  T.~Albash, V.~G.~Filev, C.~V.~Johnson and A.~Kundu,
  ``Quarks in an External Electric Field in Finite Temperature Large N Gauge
  Theory,''
  JHEP {\bf 0808}, 092 (2008),
  [arXiv:0709.1554 [hep-th]].

\bibitem{Erdmenger:2007bn}
  J.~Erdmenger, R.~Meyer and J.~P.~Shock,
  ``AdS/CFT with Flavour in Electric and Magnetic Kalb-Ramond Fields,''
  JHEP {\bf 0712}, 091 (2007),
  [arXiv:0709.1551 [hep-th]].
  
\bibitem{Karch:2007pd}
  A.~Karch and A.~O'Bannon,
  ``Metallic AdS/CFT,''
  JHEP {\bf 0709}, 024 (2007),
  [arXiv:0705.3870 [hep-th]].

\bibitem{Karch:2002sh}
  A.~Karch and E.~Katz,
  ``Adding flavor to AdS/CFT,''
  JHEP {\bf 0206}, 043 (2002),
  [arXiv:hep-th/0205236].
  
\bibitem{ft1}
We have taken the string tension to be 1 for simplicity.
  
\bibitem{Polchinski:1998rq}
For D-branes and superstring theory, see, for example,
  J.~Polchinski,
  {\sl String theory. Vol. 1, Vol. 2},
(Cambridge Univ. Press,  Cambridge 1998).

\bibitem{ft2}
The exact operator for the present model is written in Ref.~\cite{Kobayashi:2006sb}.
   
\bibitem{Kobayashi:2006sb}
  S.~Kobayashi, D.~Mateos, S.~Matsuura, R.~C.~Myers and R.~M.~Thomson,
  ``Holographic phase transitions at finite baryon density,''
  JHEP {\bf 0702}, 016 (2007),
  [arXiv:hep-th/0611099].
  
\bibitem{Nakamura:2006xk}
  S.~Nakamura, Y.~Seo, S.~J.~Sin and K.~P.~Yogendran,
  ``A new phase at finite quark density from AdS/CFT,''
  J.\ Korean Phys.\ Soc.\  {\bf 52}, 1734 (2008),
  [arXiv:hep-th/0611021].

   
\bibitem{Bergman:2008sg} 
  O.~Bergman, G.~Lifschytz and M.~Lippert,
  ``Response of Holographic QCD to Electric and Magnetic Fields,''  
  JHEP {\bf 0805}, 007 (2008),
    [arXiv:0802.3720 [hep-th]].  

\bibitem{Sonner:2012if} 
  J.~Sonner and A.~G.~Green,
  ``Hawking Radiation and Non-equilibrium Quantum Critical Current Noise,''
  Phys. Rev. Lett. {\bf 109} 091601 (2012),
  arXiv:1203.4908 [cond-mat.str-el].  

\bibitem{Alam:2012fw} 
  M.~S.~Alam, V.~S.~Kaplunovsky and A.~Kundu,
  ``Chiral Symmetry Breaking and External Fields in the Kuperstein-Sonnenschein Model,''  
  JHEP {\bf 1204}, 111 (2012),
    [arXiv:1202.3488 [hep-th]].  

\bibitem{ft3}
We write the Hamiltonian density in terms of $E$ rather than $p_{A}= \partial \widetilde{\cal L}_{\text D7}/\partial \dot{A}_{x}$ for convenience of numerical estimation.
  
\bibitem{Karch:2005ms}
  A.~Karch, A.~O'Bannon and K.~Skenderis,
  ``Holographic renormalization of probe D-branes in AdS/CFT,''
  JHEP {\bf 0604}, 015 (2006),
  [arXiv:hep-th/0512125].
  
\bibitem{Papanikolaou}
See, for example,
S.~Papanikolaou et al.,
``Universality of liquid-gas Mott transitions at finite temperatures,''
Phys. Rev. Lett. {\bf 100}, 026408 (2008),
[arXiv:0710.1627 [cond-mat.str-el]],
and references therein.

\bibitem{Limelette}
See, for example,
P.~Limelette et al.,
``Universality and Critical Behavior at the Mott transition,''
Science {\bf 302}, 89 (2003),
[arXiv:cond-mat/0406351 [cond-mat.str-el]];
F.~Kagawa, K.~Miyagawa and K.~Kanoda, 
``Unconventional critical behaviour in a quasi-two-dimensional organic conductor,''
Nature {\bf 436}, 534 (2005),
[arXiv:cond-mat/0603064 [cond-mat.str-el]].



\bibitem{CL}
See, for example,
P.~M.~Chaikin and T.~C.~Lubensky,
{\sl Principles of Condensed Matter Physics},
(Cambridge Univ. Press, Cambridge 1995).

\end{thebibliography}
\end{document}